\documentclass[prl,aps,twocolumn,twoside,superscriptaddress,notitlepage]{revtex4-1}

\usepackage{amsmath,amsfonts,amssymb,color,epsfig,graphics,graphicx,latexsym,theorem,url,verbatim}

\DeclareSymbolFont{extraup}{U}{zavm}{m}{n}
\DeclareMathSymbol{\varheart}{\mathalpha}{extraup}{86}
\DeclareMathSymbol{\vardiamond}{\mathalpha}{extraup}{87}

\newtheorem{definition}{Definition}
\newtheorem{proposition}{Proposition}
\newtheorem{lemma}{Lemma}

\newtheorem{fact}{Fact}
\newtheorem{theorem}{Theorem}
\newtheorem{corollary}{Corollary}
\newtheorem{conjecture}{Conjecture}

\newtheorem{remark}{Remark}
\newtheorem{example}{Example}
\newtheorem{question}{Question}
\newtheorem{observation}{Observation}

\def\squareforqed{\hbox{\rlap{$\sqcap$}$\sqcup$}}
\def\qed{\ifmmode\squareforqed\else{\unskip\nobreak\hfil
\penalty50\hskip1em\null\nobreak\hfil\squareforqed
\parfillskip=0pt\finalhyphendemerits=0\endgraf}\fi}
\def\endenv{\ifmmode\;\else{\unskip\nobreak\hfil
\penalty50\hskip1em\null\nobreak\hfil\;
\parfillskip=0pt\finalhyphendemerits=0\endgraf}\fi}
\newenvironment{proof}{\noindent \textbf{{Proof.~} }}{\qed}
\def\Dbar{\leavevmode\lower.6ex\hbox to 0pt
{\hskip-.23ex\accent"16\hss}D}
\makeatletter
\def\url@leostyle{%
  \@ifundefined{selectfont}{\def\UrlFont{\sf}}{\def\UrlFont{\small\ttfamily}}}
\makeatother
\def\bcj{\begin{conjecture}}
\def\ecj{\end{conjecture}}
\def\bcr{\begin{corollary}}
\def\ecr{\end{corollary}}
\def\bd{\begin{definition}}
\def\ed{\end{definition}}
\def\bea{\begin{eqnarray}}
\def\eea{\end{eqnarray}}
\def\bem{\begin{enumerate}}
\def\eem{\end{enumerate}}
\def\bex{\begin{example}}
\def\eex{\end{example}}
\def\bim{\begin{itemize}}
\def\eim{\end{itemize}}
\def\bl{\begin{lemma}}
\def\el{\end{lemma}}
\def\bpf{\begin{proof}}
\def\epf{\end{proof}}
\def\bpp{\begin{proposition}}
\def\epp{\end{proposition}}
\def\bqu{\begin{question}}
\def\equ{\end{question}}
\def\br{\begin{remark}}
\def\er{\end{remark}}
\def\bt{\begin{theorem}}
\def\et{\end{theorem}}

\def\btb{\begin{tabular}}
\def\etb{\end{tabular}}

\newcommand{\nc}{\newcommand}

 \nc{\bA}{{\bf A}} \nc{\bB}{{\bf B}}
 \nc{\bC}{{\mathbb{C}}}
 \nc{\bD}{{\bf D}} \nc{\bE}{{\bf E}} \nc{\bF}{{\bf F}}
 \nc{\bG}{{\bf G}} \nc{\bH}{{\bf H}} \nc{\bI}{{\bf I}}
 \nc{\bJ}{{\bf J}} \nc{\bK}{{\bf K}} \nc{\bL}{{\bf L}}
 \nc{\bM}{{\bf M}} \nc{\bN}{{\bf N}} \nc{\bO}{{\bf O}}
 \nc{\bP}{{\bf P}} \nc{\bQ}{{\bf Q}} \nc{\bR}{{\bf R}}
 \nc{\bS}{{\bf S}} \nc{\bT}{{\bf T}} \nc{\bU}{{\bf U}}
 \nc{\bV}{{\bf V}} \nc{\bW}{{\bf W}} \nc{\bX}{{\bf X}}
 \nc{\bZ}{{\bf Z}}

\nc{\cA}{{\cal A}} \nc{\cB}{{\cal B}} \nc{\cC}{{\cal C}}
\nc{\cD}{{\cal D}} \nc{\cE}{{\cal E}} \nc{\cF}{{\cal F}}
\nc{\cG}{{\cal G}} \nc{\cH}{{\cal H}} \nc{\cI}{{\cal I}}
\nc{\cJ}{{\cal J}} \nc{\cK}{{\cal K}} \nc{\cL}{{\cal L}}
\nc{\cM}{{\cal M}} \nc{\cN}{{\cal N}} \nc{\cO}{{\cal O}}
\nc{\cP}{{\cal P}} \nc{\cQ}{{\cal Q}} \nc{\cR}{{\cal R}}
\nc{\cS}{{\cal S}} \nc{\cT}{{\cal T}} \nc{\cU}{{\cal U}}
\nc{\cV}{{\cal V}} \nc{\cW}{{\cal W}} \nc{\cX}{{\cal X}}
\nc{\cZ}{{\cal Z}}

\nc{\hA}{{\hat{A}}} \nc{\hB}{{\hat{B}}} \nc{\hC}{{\hat{C}}}
\nc{\hD}{{\hat{D}}} \nc{\hE}{{\hat{E}}} \nc{\hF}{{\hat{F}}}
\nc{\hG}{{\hat{G}}} \nc{\hH}{{\hat{H}}} \nc{\hI}{{\hat{I}}}
\nc{\hJ}{{\hat{J}}} \nc{\hK}{{\hat{K}}} \nc{\hL}{{\hat{L}}}
\nc{\hM}{{\hat{M}}} \nc{\hN}{{\hat{N}}} \nc{\hO}{{\hat{O}}}
\nc{\hP}{{\hat{P}}} \nc{\hR}{{\hat{R}}} \nc{\hS}{{\hat{S}}}
\nc{\hT}{{\hat{T}}} \nc{\hU}{{\hat{U}}} \nc{\hV}{{\hat{V}}}
\nc{\hW}{{\hat{W}}} \nc{\hX}{{\hat{X}}} \nc{\hZ}{{\hat{Z}}}

\nc{\hn}{{\hat{n}}}

\def\tr{\mathop{\rm Tr}}
\def\w{\mathop{\rm W}}

\newcommand{\bra}[1]{\langle#1|}
\newcommand{\ket}[1]{|#1\rangle}

\def\Dbar{\leavevmode\lower.6ex\hbox to 0pt
{\hskip-.23ex\accent"16\hss}D}

\usepackage{graphicx}

\usepackage[frame,line,arrow,matrix,tips]{xy}	
\CompilePrefix{xygui-}
\makeindex

\def\w{\ar@{-}[l]}
\def\W{\ar@{=}[l]}

\def\>{\rangle}
\def\<{\langle}

\def\meter{*+[]{\put(-3,0){\includegraphics[scale=.5]{meter}}~~~~}%
		\ar@{-}[l]}

\def\gspace#1{*+{\rule[-0.2ex]{0ex}{2.1ex}%
	\setbox\sbox=\hbox{$#1$}%
	\hspace*{\wd\sbox}}}
	
\begin{document}


\title{Joint product numerical range and geometry of reduced density matrices}

\author{Jianxin Chen} %
\address{Joint Center for Quantum Information and Computer Science,
  University of Maryland, College Park, Maryland, USA}

\author{Cheng Guo} %
\address{Institute for Advanced Study, Tsinghua University, Beijing, China}

\author{Zhengfeng Ji}%
\affiliation{Centre for Quantum Computation \& Intelligent Systems,
  School of Software, Faculty of Engineering and Information
  Technology, University of Technology Sydney, Sydney, Australia}%
\affiliation{State Key Laboratory of Computer Science, Institute of
  Software, Chinese Academy of Sciences, Beijing, China}%

\author{Yiu-Tung Poon}%
\affiliation{Department of Mathematics, Iowa State University, Ames,
  Iowa, USA.}%

\author{Nengkun Yu}%
\affiliation{Institute for Quantum Computing, University of Waterloo,
  Waterloo, Ontario, Canada}%
\affiliation{Centre for Quantum Computation \& Intelligent Systems,
  School of Software, Faculty of Engineering and Information
  Technology, University of Technology Sydney, Sydney, Australia}%
\affiliation{Department of
  Mathematics \& Statistics, University of Guelph, Guelph, Ontario,
  Canada}%

\author{Bei Zeng}
\affiliation{Institute for Quantum Computing, University of Waterloo,
  Waterloo, Ontario, Canada}%
\affiliation{Department of Mathematics \& Statistics, University of
  Guelph, Guelph, Ontario, Canada}%

\author{Jie Zhou}%
\affiliation{Perimeter Institute for Theoretical Physics, Waterloo,
  Ontario, Canada}%

\begin{abstract}
The reduced density matrices of a many-body quantum
system form a convex set, whose three-dimensional projection $\Theta$
is convex in $\mathbb{R}^3$. The boundary $\partial\Theta$ of $\Theta$
may exhibit nontrivial geometry, in particular ruled surfaces.
Two physical mechanisms are known for the origins of ruled surfaces: symmetry
breaking and gapless. In this work, we study the emergence of ruled surfaces for systems
with local Hamiltonians in infinite spatial
dimension, where the reduced density matrices are
known to be separable as a consequence of the quantum de Finetti's theorem.
This allows us to identify the reduced density matrix geometry
with joint product numerical range $\Pi$ of the Hamiltonian interaction terms.
We focus on the case where the interaction terms have certain structures,
such that ruled surface emerge naturally when taking a convex hull of $\Pi$. We show that,
a ruled surface on $\partial\Theta$ sitting in $\Pi$ has a gapless origin, otherwise it has
a symmetry breaking origin. As an example, we demonstrate that a famous ruled surface,
known as the oloid, is a possible shape of $\Theta$, with two boundary pieces of
symmetry breaking origin separated by two gapless lines.
\end{abstract}


\maketitle

\section{I. Introduction}

In a natural many-body quantum system, the Hamiltonian $H$ usually involves only
two-body interactions. Consequently, for any many-body wave function $\ket{\psi}$,
its energy $E_{\psi}=\bra{\psi}H\ket{\psi}$ only depends on the two-particle reduced density
matrix ($2$-RDM) of $\ket{\psi}$. In case $H$ depends on some parameters $\vec{\lambda}$, the ground state energy $E_0(\vec{\lambda})$ of the system may exhibit non-analylic behaviour while $\vec{\lambda}$ change smoothly, where a quantum phase transition occurs. Since the energy $E_{\psi}$ only depends on its $2$-RDMs, it is much desired that the geometry
of the $2$-RDMs may capture such a sudden change in ground state energy $E_0(\vec{\lambda})$.

The set of all $2$-RDMs is known to be convex, however its shape is notoriously difficult to characterize in general. Since 1960s, how to characterize this convex set has been a central topic of research in the field of quantum marginal problem and $N$-representability problem~\cite{Col63,Erd72,klyachko2006quantum,EJ00,SM09}. The recent development in quantum information theory has shown that the characterization of
the $2$-RDMs is a hard problem even with the existence of a quantum
computer~\cite{Liu06,LCV07,WMN10}. Nevertheless, many practical
approaches are developed to characterize the properties of $2$-RDMs,
and to retrieve useful information that reflects the physical
properties of the system~\cite{VC06,GM06}.

One important idea is to study the properties of $2$-RDMs is
by looking at the two- and three-
dimensional projections~\cite{EJ00,VC06,GM06,SM09}. Since these projections
are convex sets in $\mathbb{R}^2$ and $\mathbb{R}^3$ respectively,
the hope is that the properties of the different quantum phases can be visually
available. Interestingly, it has been shown that a flat portion of the
two-dimensional projection can already signal first-order phase
transitions~\cite{chen2015discontinuity,zauner2014symmetry}.
However, for continuous phase transitions, two-dimensional projections contain
no information, and one needs to look further at the three-dimensional
projections.

In~\cite{zauner2014symmetry} it is observed that the emergence of ruled surfaces on the boundary of
the three-dimensional projections of $2$-RDMs can be a signal of
symmetry breaking phase. With a
generalization to non-thermodynamic observables, the ruled surfaces can also signal the symmetry
protected topological phase~\cite{chen2016geometry}. Very recently, it
is also observed that gapless systems can also lead to ruled surfaces,
and two examples of such systems are discussed, which are both interacting
many-body bosonic systems~\cite{chen2016physical}.

It was Gibbs in 1870's who first proposed the deep
connection between ruled surfaces on the boundary of certain convex body
and phase transitions~\cite{Gib73a,Gib73b,Gib75,Isr79}, in the
context of classical thermodynamics, which reflects
a fundamental property of thermodynamic stability.
Although the convex set considered here is in terms of quantum
many-body physics, and the connection is between
ruled surface on the projection of RDMs and quantum phase transitions,
it nevertheless indicates that the convex geometry approach is 
a fundamental and universal idea.

In both classical and quantum cases,
one fundamental question is in the reverse direction: that is,
what kind of ruled surfaces are actually possible? In other words,
what shapes of ruled surfaces may actually correspond to a practical
quantum system? This question is, of course, too hard in general, but opens
some interesting possibilities. For instance, one may ask whether
the oloid, being a famous ruled surface in $\mathbb{R}^3$,
can be a three-dimensional
projection of some convex set of $2$-RDMs.

This oloid idea may sound unrealistic at the first sight. Surprisingly,
we will show that this is in fact possible. We will develop a method
that systematically leads to many other possibilities of ruled surface for
the three-dimensional projections of
$2$-RDMs. We start from a fact that although 
the geometry of $2$-RDMs are in general hard to characterize,
there is one situation it is provably easy: that is, for an infinite spatial dimensional
system, the $2$-RDMs are known to be separable, due to the celebrated
quantum de Finetti's theorem~\cite{stormer1969symmetric,hudson1976locally,lewin2014derivation}.

This simplification to only separable states
then allows us to study the geometry of $2$-RDMs
with a mathematical concept, called joint product numerical range~\cite{puchala2011product,
dirr2008relative,duan2008local,schulte2008significance,schulte2010gradient,gawron2010restricted},
denoted by $\Pi$, of the Hamiltonian interaction terms.
$\Pi$ includes all the extreme points of the three-dimensional projections of
$2$-RDMs, and the projection itself,
denoted by $\Theta$, is a convex hull of $\Pi$.
We then focus on the cases where the interaction terms have certain structures,
such that ruled surfaces emerge naturally when taking a convex hull of $\Pi$. We show that,
for a ruled surface on the boundary of $\Theta$, denoted by $\partial\Theta$, if it also sits in $\Pi$, then it has a gapless origin, otherwise it has a symmetry breaking origin.

\section{II. Reduced density matrix geometry and its projections}

We consider many-body systems with $N$-particles, and single-particle dimension $d$ (i.e.
single-particle Hilbert space $\mathbb{C}^d$).
For a many-body Hamiltonian $H(\vec{\lambda})$, we discuss the case where
$\vec{\lambda}=(\lambda_1,\lambda_2,\lambda_3)$, and
\begin{equation}
H(\vec{\lambda})=\sum_{i=1}^3 \lambda_i H_i.
\end{equation}
Here each $H_i=\sum_{j} h_{j,i}$, and $h_{j,i}$ involves at most two-body interactions. Therefore,
for any many-body wave function of the system, only its $2$-RDM is of relevance to our discussion.
In other words, we are interested in the set of all possible $2$-RDMs of the many-body wave
functions.

In practice, the structure of the $2$-RDMs only depends on the interaction pattern of
$H$~\cite{zeng2015quantum}. That is, usually, the interaction terms in $H_i$ involves only `nearest-neighbour'
interactions depending on the spatial geometry of the system. In this work we consider
a special case, where the spatial geometry is infinite-dimensional. That is, each
single particle in the system has infinitely-many neighbours. Consequently, we are in
the limit of infinite number of particles, i.e. $N\rightarrow\infty$.

For simplicity we consider a particular case that the system has $\mathcal{A}$, $\mathcal{B}$ sub-lattices and with translational symmetry, and the $H_i$\,s only involve nearest-neighbour
interactions. That is,  for $H_i=\sum_{j}h_{j,i}$, each $h_{j,i}$ involves
at most two-body interactions and acts the same for each nearest neighbour $AB$ particles.
And for any particle $A$ in the sublattice $\mathcal{A}$ and particle $B$ in the sublattice $\mathcal{B}$
that are neighbours, the corresponding reduced density matrix $\rho_{AB}$ (of any state of the $N$-particle system) are the same. In other words, $\rho_{AB}$ contains
all the information of interest of $2$-RDMs of the physical system.

In the $N\rightarrow\infty$ limit, the quantum de Finetti's theorem guarantees that
$\rho_{AB}$ is separable~\cite{stormer1969symmetric,hudson1976locally}. In other words,
$\rho_{AB}$ can always be written as
\begin{equation}
\rho_{AB}=\sum_j c_j\ket{\psi^j_{AB}}\bra{\psi^j_{AB}}\,,
\end{equation}
with $c_j\geq 0,\ \sum_j c_j=1$, and each $\ket{\psi^j_{AB}}\in\mathbb{C}^d\otimes\mathbb{C}^d$ is
a product state of the form
\begin{equation}
\ket{\psi^j_{AB}}=\ket{\alpha^j}\otimes\ket{\beta^j}\,,
\end{equation}
with $\ket{\alpha^j},\ket{\beta^j}\in\mathbb{C}^d$.

Therefore, to study the three-dimensional projection of the $2$-RMDs, we
are in fact considering the three-dimensional projection of the set of
all the two-particle separable state $\rho_{AB}$s. This projection is
given by the set of points $(x,y,z)\in\mathbb{R}^3$, where
\begin{equation}
x=\tr (H_1\rho_{AB}),\ y=\tr(H_2\rho_{AB}),\ z=\tr(H_3\rho_{AB}),
\end{equation}
And the projection of the extreme points of the set of all separable state $\rho_{AB}$s, which are product states $\ket{\alpha}\otimes\ket{\beta}$, is given by the set of points $(x,y,z)\in\mathbb{R}^3$, where
\begin{eqnarray}
\label{eq:prod}
x&=&(\bra{\alpha}\otimes\bra{\beta})H_1(\ket{\alpha}\otimes\ket{\beta}),\nonumber\\
y&=&(\bra{\alpha}\otimes\bra{\beta})H_2(\ket{\alpha}\otimes\ket{\beta}),\nonumber\\
z&=&(\bra{\alpha}\otimes\bra{\beta})H_3(\ket{\alpha}\otimes\ket{\beta}).
\end{eqnarray}

And in fact we only need to consider the terms $h_{j,i}$ of $H_i$ that act non-trivially on particles $AB$.
In other words, we can equivalently consider $H_i$ as Hermitian operators acting on $\mathbb{C}^d\otimes\mathbb{C}^d$, without confusion we use $H_i$  to mean its `energy per particle' version acting on 
two particles $AB$ (hence $H_i$ is bounded)~\cite{chen2016physical}. This then allows us to connect our discussions to some mathematical concepts, namely certain kind of joint numerical ranges of
$H_i$\,s. For simplicity we only consider $d=2$ (i.e. qubit) case in this work. However, the method we discuss is general and can extend to the $d>2$ cases.

\section{III. Product numerical range}

We consider a two-qubit system $AB$, with Hilbert space $\mathbb{C}^2\otimes\mathbb{C}^2$. Let $S $ be the set of normalized $\ket{\psi}\in\mathbb{C}^2\otimes\mathbb{C}^2$ (i.e. $\bra{\psi}{\psi}\rangle=1$).
For any three $4\times 4$ Hermitian operators $H_1,H_2,H_3$,
the joint numerical range~\cite{AP} of $H_1,H_2,H_3$ is given by
\begin{eqnarray}
&&\Lambda(H_1,H_2,H_3)=\nonumber\\
&&\{(\bra{\psi}H_1\ket{\psi},\bra{\psi}H_2\ket{\psi},\bra{\psi}H_3\ket{\psi})
|\ket{\psi}\in S\}\,.
\end{eqnarray}

One important property of $\Lambda(H_1,H_2,H_3)$ is given in~\cite{AP} that is summarized below.
\begin{fact}
$\Lambda(H_1,H_2,H_3)\subset\mathbb{R}^3$ is convex.
\end{fact}

Let $S_{\Pi}$ be the set of product states $\ket{\phi}=\ket{\alpha}\otimes\ket{\beta} \in S$ with $\ket{\alpha},\ket{\beta}\in\mathbb{C}^2$, the product numerical range of $H_1,H_2,H_3$ is given by
\begin{eqnarray}
&&\Pi(H_1,H_2,H_3)=\nonumber\\
&&\{(\bra{\phi}H_1\ket{\phi},\bra{\phi}H_2\ket{\phi},\bra{\phi}H_3\ket{\phi})
|\ket{\phi}\in S_{\Pi}\}.
\end{eqnarray}
It is known that $\Pi(H_1,H_2,H_3)$ is in general not convex~\cite{PG}.

Consider any separable state
$
\rho_{AB}=\sum_j c_j\ket{\psi^j_{AB}}\bra{\psi^j_{AB}}
$
with each $\ket{\psi^j_{AB}}=\ket{\alpha^j}\otimes\ket{\beta^j}$ a product state. The separable numerical range of $H_1,H_2,H_3$ is given by
\begin{eqnarray}
&&\Theta(H_1,H_2,H_3)=\nonumber\\
&&\{(\tr H_1\rho_{AB},\tr H_2\rho_{AB},\tr H_3\rho_{AB})\nonumber\\
&&|\rho_{AB}\ \text{separable}\}.
\end{eqnarray}
It is clear that 
$\Theta(H_1,H_2,H_3)$ is the convex hull of $\Pi(H_1,H_2,H_3)$, hence is convex,
with all the extreme points in $\Pi(H_1,H_2,H_3)$. In general
\begin{eqnarray}
\label{eq:ThetaLamda}
\Theta(H_1,H_2,H_3)\subseteq\Lambda(H_1,H_2,H_3),
\end{eqnarray}
and in most cases, $\Theta(H_1,H_2,H_3)$ does not equal to $\Lambda(H_1,H_2,H_3)$.

\subsection{A. The physical origin of boundary ruled surfaces}
\label{sec:phy}

For any product state $\ket{\psi}=\ket{\alpha}\otimes\ket{\beta}$ and the Hamiltonian $H(\vec{\lambda})=\sum_{i=1}^3\lambda_iH_i$, 
its energy is
\begin{equation}
E_{\psi}(\vec{\lambda})=x\lambda_1+y\lambda_2+z\lambda_3\geq E_0(\vec{\lambda}),
\end{equation}
where $x,y,z$ are given in Eq.~\eqref{eq:prod}
and $E_0(\vec{\lambda})$ is the ground state energy of $H(\vec{\lambda})$. 
Since $\Theta(H_1,H_2,H_3)$ is convex, for each $\vec{\lambda}$,
the Hamiltonian $H(\vec{\lambda})$ can be interpreted 
as a supporting plane of $\Theta(H_1,H_2,H_3)$ with normal 
vector given by $\vec{\lambda}=(\lambda_1,\lambda_2,\lambda_3)$~\cite{chen2012ground,chen2015discontinuity}. 

Our main focus is on the boundary of $\Theta(H_1,H_2,H_3)$,
which is denoted by $\partial\Theta(H_1,H_2,H_3)$. 
Generically,
an exposed point $P_e$ on $\partial\Theta(H_1,H_2,H_3)$ has a unique
product state pre-image in $\Pi(H_1,H_2,H_3)$. Physically, this means
that the corresponding Hamiltonian $H(\vec{\lambda})=\sum_{i=1}^3 \lambda_i H_i$ (i.e.
the supporting plane of $\Theta(H_1,H_2,H_3)$ that intersects
$\partial\Theta(H_1,H_2,H_3)$ at the point $P_e$) has a unique product ground state. 

The boundary of $\partial\Theta(H_1,H_2,H_3)$ can also be flat.
And in most cases this plat portion 
is completely flat, i.e. it is a part of a plane, which is
an area of the intersection of the corresponding supporting plane of $\Theta(H_1,H_2,H_3)$
with $\partial\Theta(H_1,H_2,H_3)$. The boundary of the area
has infinitely many product state pre-images in $\Pi(H_1,H_2,H_3)$.
Physically, this means that the corresponding Hamiltonian is gapless.

A nontrivial case is that the flat portion on $\partial\Theta(H_1,H_2,H_3)$
is not a part of a plane, but rather a
ruled surface. That is, for any point $P_e$ on $\partial\Theta(H_1,H_2,H_3)$, there is a line segment
$L$ passing $P_e$ that is also on the surface.
Physically, there are two known origins
of ruled surfaces: 1) symmetry breaking~\cite{zauner2014symmetry} and
2) Gapless~\cite{chen2016physical}.

In general, one cannot tell the physical
origins of the ruled surfaces solely from the shape of
$\partial\Theta(H_1,H_2,H_3)$~\cite{chen2016physical}.
One idea to tell the difference is to look at the finite scaling of
RDM geometry~\cite{chen2016physical}. Here we
would like to connect these physical origins to the properties of
joint product numerical range $\Pi(H_1,H_2,H_3)$.

Consider any line segment $L$ on $\partial\Theta(H_1,H_2,H_3)$,
with two end points $P_a$ and $P_b$. In case there is no plane area
on $\partial\Theta(H_1,H_2,H_3)$ that contains $L$, then there is 
a supporting plane of $\Theta(H_1,H_2,H_3)$ that intersects
$\partial\Theta(H_1,H_2,H_3)$ only at $L$, with normal vector
$\vec{\lambda}=(\lambda_1,\lambda_2,\lambda_3)$.

It is clear that $P_a,P_b\in\Pi(H_1,H_2,H_3)$.
Here are two possible cases: 1) $L$ is not in $\Pi(H_1,H_2,H_3)$,
2) $L\subset\Pi(H_1,H_2,H_3)$. For case 1), generically, each $P_a$
(or $P_b$) has a unique product state pre-image in $\Pi(H_1,H_2,H_3)$.
Consequently, the corresponding Hamiltonian 
$H=\sum_{i=1}^3 \lambda_i H_i$ has degenerate product ground states.
For case 2), each 
point on $L$ has a product state pre-image in $\Pi(H_1,H_2,H_3)$,
so $L$ has infinitely many product state pre-images in $\Pi(H_1,H_2,H_3)$.
Consequently, the corresponding Hamiltonian 
$H=\sum_{i=1}^3 \lambda_i H_i$ is gapless.

If there is a piece of ruled surface on $\partial\Theta(H_1,H_2,H_3)$ which
is not in $\Pi(H_1,H_2,H_3)$ (except the boundary of the piece), then 
the corresponding Hamiltonian 
$H=\sum_{i=1}^3 \lambda_i H_i$ maintains
its round state degeneracy when $\vec{\lambda}$ varies. In other words, for
a range of parameters $\vec{\lambda}$, $H=\sum_{i=1}^3 \lambda_i H_i$
has stable ground state degeneracy (along the parameter-changing direction),
with product ground states. This is a typical feature of symmetry breaking. In this sense,
a ruled surface on $\partial\Theta(H_1,H_2,H_3)$ that is not 
in $\Pi(H_1,H_2,H_3)$ has a symmetry breaking origin. In comparison,
if the ruled surface piece is in $\Pi(H_1,H_2,H_3)$, then it has 
a gapless origin.

We summarize our observation as the following.
\begin{observation}
A ruled surface on
\begin{equation}
\Pi(H_1,H_2,H_3)\cap\partial\Theta(H_1,H_2,H_3)
\end{equation}
has a gapless origin. Otherwise, a ruled surface on $\partial\Theta(H_1,H_2,H_3)$ that
is not in $\Pi(H_1,H_2,H_3)$ has a symmetry breaking origin.
\end{observation}

This observation gives us a general method to study the physics of 
the reduced density matrix geometry 
$\Theta(H_1,H_2,H_3)$. That is, for any given system $H=\sum_{i=1}^3 \lambda_i H_i$,
by comparing $\Theta(H_1,H_2,H_3)$
with $\Pi(H_1,H_2,H_3)$, one should be able to get quantum phases and phase transition
informations by solely looking at the geometry of  $\Theta(H_1,H_2,H_3)$ and $\Pi(H_1,H_2,H_3)$.

\subsection{B. The block diagonal Hamiltonians}
\label{sec:block}

Compared to exposed points and completely flat areas, ruled surfaces are 
much less possible to find in generic systems. It usually requires certain structure of $H=\sum_{i=1}^3 \lambda_i H_i$, e.g. symmetry. In order to understand the possible ruled surface shapes on $\partial\Theta(H_1,H_2,H_3)$
and their physical possible origins, we would like to look at Hamiltonians with structure. Geometrically,
flat portions on  $\partial\Theta(H_1,H_2,H_3)$ may be obtained from the convex hull of two (not flat)
objects. This inspires us to consider the case where $H_1,H_2,H_3$ are block diagonal, i.e.
\begin{equation}
H_i=
\begin{pmatrix}
H_i^a & O \\
O & H^b_i
\end{pmatrix},\ i=1,2,3\,,
\end{equation}
where $H_i^a, H_i^b$ are $2\times 2$, and $O$ is the $2\times 2$ zero matrix.

Denote the joint numerical range of $H_1^a,H_2^a,H_3^a$ by $\Lambda(H^a_1,H_2^a,H_3^a)$ and
the joint numerical range of $H_1^b,H_2^b,H_3^b$ by $\Lambda(H^b_1,H_2^b,H_3^b)$. One important
property of joint numerical range of block diagonal matrices is the following.

\begin{fact}
$\Lambda(H_1,H_2,H_3)$ is the convex hull of $\Lambda(H^a_1,H_2^a,H_3^a)$ and $\Lambda(H^b_1,H_2^b,H_3^b)$~\cite{BL}.
\end{fact}

What we need is the shape of $\Theta(H_1,H_2,H_3)$, not $\Lambda(H_1,H_2,H_3)$. They
are in general two very different sets, i.e. the equality in Eq.~\eqref{eq:ThetaLamda} in general does not
hold. In the special case of block diagonal matrices, however, 
we can show that these two sets coincide.

\begin{observation}
\label{obs:main}
For block diagonal $H_1,H_2,H_3$,
\begin{equation}
\Theta(H_1,H_2,H_3)=\Lambda(H_1,H_2,H_3)
\end{equation}
\end{observation}

To show why it is the case, let
$
H_i=
\begin{pmatrix}
H_i^a & O \\
O & H^b_i
\end{pmatrix},$
for $i=1,2,3$. Suppose $|\alpha\rangle\in \bC^2$. Then
\begin{equation}
\langle \alpha|H^a_i|\alpha\rangle=(\langle 0|\otimes\langle \alpha|)H_i(|0\rangle \otimes|\alpha\rangle) \,.
\end{equation}
Similarly,
\begin{equation}
\langle \alpha|H^b_i|\alpha\rangle=(\langle 1|\otimes\langle \alpha|)H_i(|1\rangle \otimes|\alpha\rangle) \,.
\end{equation}
Therefore,
\begin{equation}
\Lambda(H^a_1, H^a_2, H^a_3) ,\ \Lambda(H^b_1, H^b_2, H^b_3)\subseteq \Pi(H_1, H_2, H_3)\,.
\end{equation}

Since
\begin{eqnarray}
&&\Theta(H_1, H_2, H_3) \nonumber\\
=&&\mbox{  conv }\Pi (H_1, H_2, H_3)\subseteq \Lambda(H_1, H_2, H_3) \nonumber\\
=&&\mbox{  conv }\{\Lambda(H^a_1, H^a_2, H^a_3) ,\ \Lambda(H^b_1, H^b_2, H^b_3) \}\,,
\end{eqnarray}
we have
$
\Theta(H_1, H_2, H_3)=\Lambda(H_1, H_2, H_3)\,.
$

\subsection{C. The oloid}

\begin{figure}[htbp]
  \centering
  \includegraphics[width=2.5in]{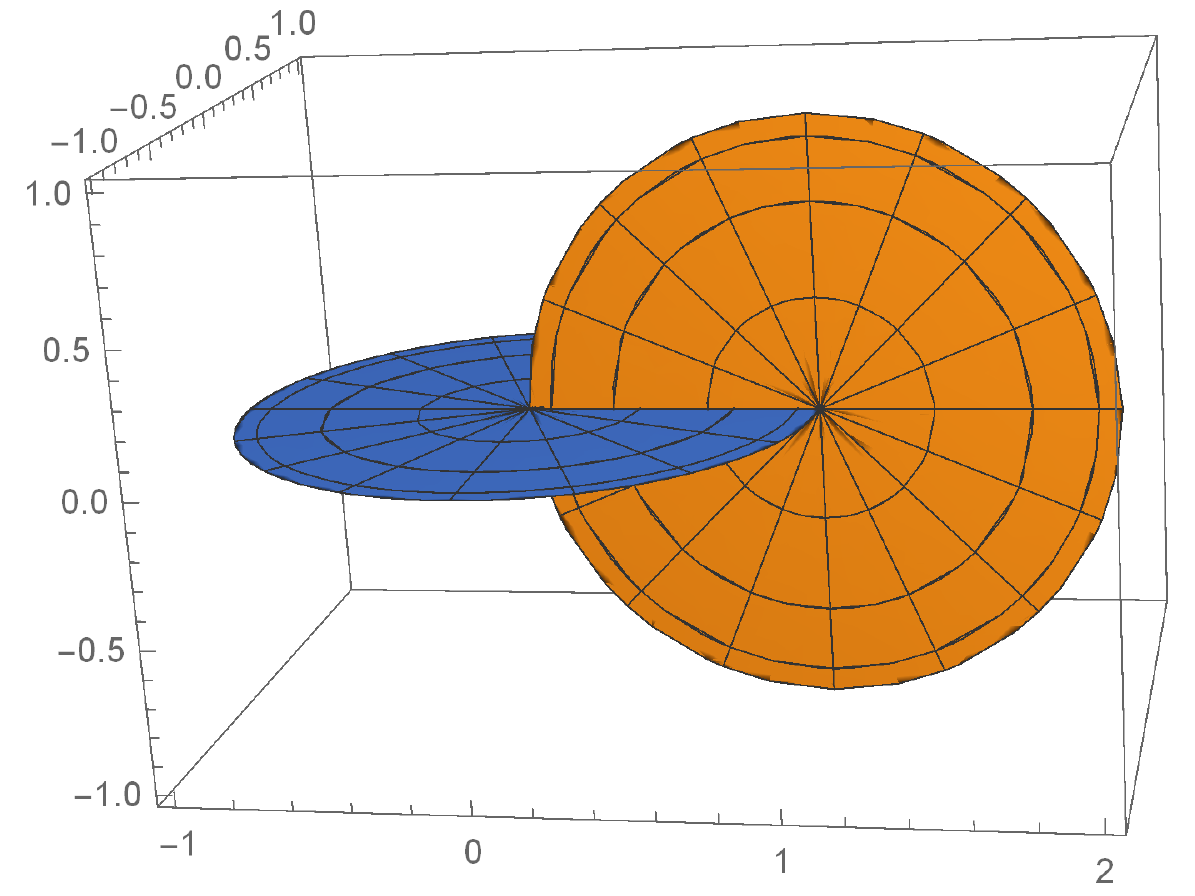}
  \caption{The two disks corresponding to $\Lambda(H_1^a, H_2^a, H_3^a)$ and
  $\Lambda(H_1^b, H_2^b, H_3^b)$.}
  \label{fig:2disk}
\end{figure}

As an example, let
\begin{eqnarray}
H_1^a=\begin{pmatrix} 0&1\\ 1&0\end{pmatrix},
\ H_2^a=\begin{pmatrix} 0&-i\\ i&0\end{pmatrix},
\ H_3^a=\begin{pmatrix} 0&0\\ 0&0\end{pmatrix}\,,
\end{eqnarray}
then $\Lambda(H_1^a, H_2^a, H_3^a)$ is the disk
$\{(x,y,0)\in \mathbb{R}^3:x^2+y^2\le 1\}$. Similarly, if
\begin{eqnarray}
H_1^b=\begin{pmatrix} 1&1\\ 1&1\end{pmatrix},
\ H_2^b=\begin{pmatrix} 0&0\\ 0&0\end{pmatrix},
\ H_3^b=\begin{pmatrix} 0&-i\\ i&0\end{pmatrix}\,,
\end{eqnarray}
then
$\Lambda(H_1^b, H_2^b, H_3^b)$ is the
disk $\{(x,0,z)\in \mathbb{R}^3:(x-1)^2+z^2\le 1\}$. 
A plot of these two disks is given in Fig.~\ref{fig:2disk}.

\begin{figure}[htbp]
  \centering
  \includegraphics[width=1.5in,angle=-90]{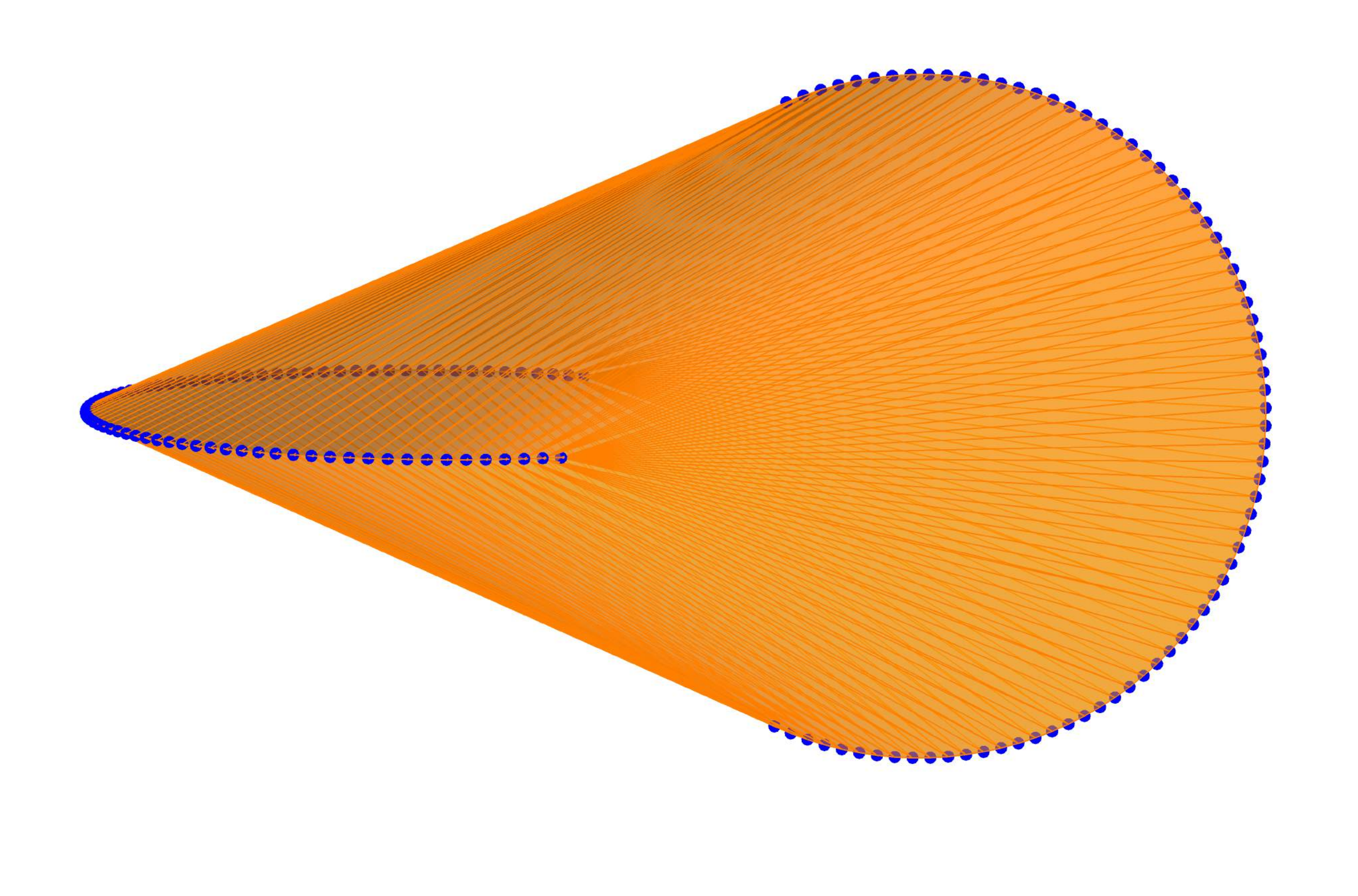}
  \caption{An Oloid.}
  \label{fig:oloid}
\end{figure}

Therefore,
\begin{eqnarray}
&&\Theta(H_1, H_2, H_3)=\nonumber\\
&&\mbox{  conv }\{\Lambda(H^a_1, H^a_2, H^a_3) ,\ \Lambda(H^b_1, H^b_2, H^b_3) \}
\end{eqnarray}
is the so called `oloid'~\cite{dirnbock1997development}. A illustration of the oloid is given in Fig.~\ref{fig:oloid}.

To obtain the joint product numerical range $\Pi(H_1, H_2, H_3)$,
suppose
\begin{eqnarray}
|\alpha\rangle =u_1\ket{0}+u_2\ket{1}
\end{eqnarray}
with $|u_1|^2+|u_2|^2=1$,
and
\begin{eqnarray}
|\beta\rangle =r\ket{0}+\sqrt{1-r^2}e^{it}\ket{1}
\end{eqnarray} 
with $0\leq r\leq 1$ and $0\leq t\leq 2\pi$. Then
\begin{eqnarray}
\begin{array}{rl}&(\langle \alpha| \otimes \langle \beta|)(H_1, H_2, H_3)(|\alpha \rangle\otimes |\beta \rangle ) \\&\\ =&|u_1|^2(2r\sqrt{1-r^2}\cos t,2r\sqrt{1-r^2}\sin t , 0)\\&\\ +&
|u_2|^2(1+2r\sqrt{1-r^2}\cos t,0,2r\sqrt{1-r^2}\sin t)\end{array}
\end{eqnarray} 

Therefore, let $|u_1|^2=s_1$, $2r\sqrt{1-r^2}=s_2$, we have

\begin{eqnarray}
&&\Pi(H_1, H_2, H_3)=\nonumber\\
&&\{(1-s_1+s_2\cos t,s_1s_2\sin t,(1-s_1)s_2\sin t):\nonumber\\
&& 0\le s_1,\ s_2\le 1,\ 0\le t \le 2\pi\}\,.
\end{eqnarray}

\begin{figure}[htbp]
  \centering
  \includegraphics[width=2.5in]{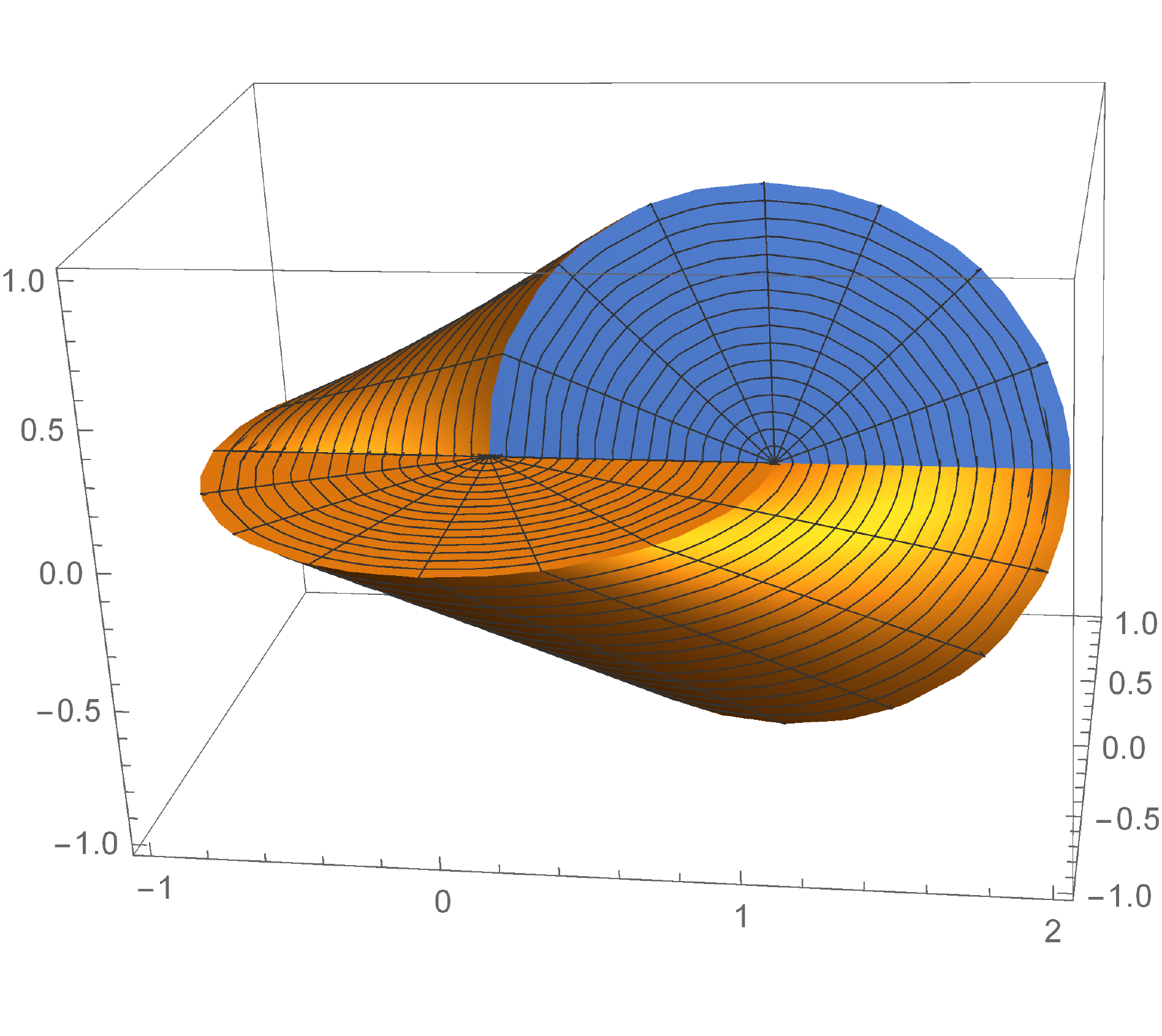}
  \caption{$\Pi(H_1, H_2, H_3)$ for oloid.}
  \label{fig:oloid1}
\end{figure}

We illustrate $\Pi(H_1, H_2, H_3)$ in Fig.~\ref{fig:oloid1}.

\begin{figure}[htbp]
  \centering
  \includegraphics[width=2.0in]{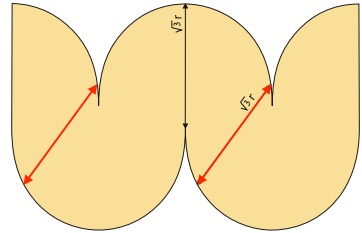}
  \caption{The developable surface of the oloid. The two red lines
  are in $\Pi(H_1, H_2, H_3)\cap\partial\Theta(H_1, H_2, H_3)$. Figure modified from https://en.wikipedia.org/wiki/Oloid. }
  \label{fig:oloiddep}
\end{figure}

We are interested in the shape of 
$\Pi(H_1, H_2, H_3)\cap\partial\Theta(H_1, H_2, H_3)$.
Since the oloid is a developable surface, we can expand its
boundary to put on a plane, as shown in Fig.~\ref{fig:oloiddep}.
The intersection of $\Pi(H_1, H_2, H_3)$ with $\partial\Theta(H_1, H_2, H_3)$
contains all the curved parts of the boundary (i.e. boundaries of
the two disks). In addition, it also contains
two lines shown as the red lines in Fig.~\ref{fig:oloiddep}. These two red lines
cut $\partial\Theta(H_1, H_2, H_3)$ into two pieces of ruled surfaces, each of which corresponds
to a symmetry breaking phase as discussed in Sec.~IIIA. That is,
the physical system $H(\vec{\lambda})=\sum_{i=1}^3 \lambda_iH_i$ has two 
symmetry breaking phases separated by two gapless transition points (that correspond
to the two red lines in Fig.~\ref{fig:oloiddep}).

\section{IV. The symmetric case and bosonic systems}
\label{sec:boson}

We can also consider the product numerical range in the symmetric case. That is, instead of considering
product state of the form $\ket{\alpha}\otimes\ket{\beta}$, we restrict ourselves in the case of $\ket{\alpha}\otimes\ket{\alpha}$,
and the $H_i$\,s are supported on the symmetric subspace of $\mathbb{C}^2\otimes\mathbb{C}^2$.
For the symmetric case, we denote the corresponding joint product numerical range by
$\Pi_{+}(H_1,H_2,H_3)$, and the joint separable numerical range by $\Theta_{+}(H_1,H_2,H_3)$.
Physically, we are dealing with a many-body bosonic system with symmetric wavefunctions in the $N\rightarrow\infty$ limit, 
where the reduced density matrices of the wave function of the system is also known to be separable
due to the quantum de Finetti's theorem~\cite{lewin2014derivation}.

We consider $\Pi_{+}(H_1,H_2,H_3)$ that is given by the set of points $(x,y,z)\in\mathbb{R}^3$, where
\begin{eqnarray}
\label{eq:xyz}
  x&=&\bra{\alpha^{\otimes 2}}H_1\ket{\alpha^{\otimes 2}}=\tr(H_1\ket{\alpha^{\otimes 2}}\bra{\alpha^{\otimes 2}}),\nonumber\\
  y&=&\bra{\alpha^{\otimes 2}}H_2\ket{\alpha^{\otimes 2}}=\tr(H_2\ket{\alpha^{\otimes 2}}\bra{\alpha^{\otimes 2}}),\nonumber\\
  z&=&\bra{\alpha^{\otimes 2}}H_3\ket{\alpha^{\otimes 2}}=\tr(H_3\ket{\alpha^{\otimes 2}}\bra{\alpha^{\otimes 2}})\,.
\end{eqnarray}
Here $\ket{\alpha}\in\mathbb{C}^2$ is any single qubit state.

We can parameterize
\begin{equation}
\ket{\alpha}\bra{\alpha}=\frac{1}{2}(I+rX+sY+tZ)\,,
\end{equation}
and $\vec{r}=(r,s,t)^T$, with $\vec{r}^{T}\vec{r}=1$.

And each $H_i$ can be written in the Pauli basis as
\begin{eqnarray}
&&H_i=c_{0,i}+c_{xx,i}X\otimes X+c_{yy,i}Y\otimes Y+ c_{zz,i} Z\otimes Z\nonumber\\
&&+c_{xy,i}(X\otimes Y+Y\otimes X)+c_{x,i}(X\otimes I+I\otimes X)\nonumber\\
&&+c_{yz,i}(Y\otimes Z+Z\otimes Y)+c_{y,i}(Y\otimes I+I\otimes Y)\nonumber\\
&&+c_{xz,i}(X\otimes Z+Z\otimes X)+c_{z,i}(Z\otimes I+I\otimes Z)
\end{eqnarray}

With this parameterization, our joint product numerical range of $H_1,H_2,H_3$ becomes the set of points in $\mathbb{R}^3$ given by
\begin{equation}
\label{eq:fi}
\left(f_1(r,s,t), f_2(r,s,t),f_3(r,s,t)\right),
\end{equation}
where each $f_i$ is a polynomial of $r,s,t$ of degree at most $2$, with the constraint
$
r^2+s^2+t^2=1.
$

\subsection{A. The homogenous case}

We first consider the simple case where the $f_i$\,s are homogenous polynomials
of $r,s,t$, i.e. $c_{0,i}=c_{x,i}=c_{y,i}=c_{z,i}=0$. Hence we can rewrite
\begin{eqnarray}
\label{eq:realNR}
  x=\vec{r}^{T}M_1\vec{r},\ 
  y=\vec{r}^{T}M_2\vec{r},\ 
  z=\vec{r}^{T}M_3\vec{r},
\end{eqnarray}
where 
\begin{equation}
M_i=
\begin{pmatrix}
c_{xx,i} & d_{xy,i} & d_{xz,i}\\
d_{xy,i} & c_{yy,i} & d_{yz,i}\\
d_{xz,i} & d_{yz,i} & c_{zz,i}
\end{pmatrix}
\end{equation}
is a real symmetric matrix, for $i=1,2,3$.

Now consider the operator
\begin{equation}
M=uM_1+vM_2+wM_3\,,
\end{equation}
which is a real symmetric matrix,
whose eigenvectors can be all real.

Notice that
\begin{equation}
\Pi_{+}(H_1,H_2,H_3)=\{(\vec{r}^{T}M_1\vec{r},\vec{r}^{T}M_2\vec{r},\vec{r}^{T}M_3\vec{r})\}\,,
\end{equation}
for all $\vec{r}^{T}\vec{r}=1$,
is the real version of the joint numerical range of $(M_1,M_2,M_3)$. To compare with
the joint numerical range of $(M_1,M_2,M_3)$, we denote
\begin{equation}
\Lambda_{\mathbb{R}}(M_1,M_2,M_3)=\{(\vec{r}^{T}M_1\vec{r},\vec{r}^{T}M_2\vec{r},\vec{r}^{T}M_3\vec{r})\}.
\end{equation}

Although $\Lambda_{\mathbb{R}}(M_1,M_2,M_3)$ may not be convex~\cite{AP}, we are going to show that every point in $\Lambda(M_1,M_2,M_3)$ is a convex combination of (at most) two points in $\Lambda_{\mathbb{R}}(M_1,M_2,M_3)$. Therefore, the extreme points of $\Lambda(M_1,M_2,M_3)$ lie in $\Lambda_{\mathbb{R}}(M_1,M_2,M_3)$.

Suppose $M$ is real and symmetric $n\times n$ matrix,
and $\ket{v} \in \bC^n$ satisfies $\bra{v}{v}\rangle=1$.  
Let 
\begin{equation}
\ket{v} =\ket{x}+i\ket{y}\,, 
\end{equation}
where $\ket{x},\ \ket{y}  \in \mathbb{R}^n$. 
Then 
\begin{equation}
\bra{v}{v}\rangle=1\Rightarrow \bra{x}{x}\rangle+\bra{y}{y}\rangle=1\,.
\end{equation}
We have
\begin{eqnarray}
\bra{v}M\ket{v}&=&(\bra{x}-i\bra{y})M(\ket{x}+i\ket{y})\nonumber\\
&=&\bra{x}M\ket{x}+\bra{y}M\ket{y}\,.
\end{eqnarray}
If $\ket{x}$ or $\ket{y}$ is the zero vector, then 
\begin{eqnarray}
\bra{v}M\ket{v}\in  \Lambda_{\mathbb{R}}(M_1,M_2,M_3). 
\end{eqnarray}
Suppose both $\ket{x}$ and $\ket{y}$ are nonzero. 
Let $t=\sqrt{\bra{x}{x}\rangle}$, and 
\begin{eqnarray*}
\ket{x'}=\frac{1}{t}\ket{x},\ \text{and}\ \ket{y'}=\frac{1}{\sqrt{1-t^2}}\ket{y}. 
\end{eqnarray*}
Then $\bra{x'}{x'}\rangle=\bra{y'}{y'}\rangle=1$ and
\begin{equation}
\bra{v}M\ket{v}=t\bra{x'}M\ket{x'}+(1-t)\bra{y'}M\ket{y'}\,.
\end{equation}

That is, the extreme points of $\Lambda(M_1,M_2,M_3)$ lie in $\Lambda_{\mathbb{R}}(M_1,M_2,M_3)=\Pi_{+}(H_1,H_2,H_3)$.
Since $\Lambda(M_1,M_2,M_3)$ is convex and  the convex hull of $\Pi_{+}(H_1,H_2,H_3)$ is  $\Theta_+(H_1,H_2,H_3)$, we have
\begin{observation}
\begin{eqnarray}
\Theta_+(H_1,H_2,H_3)=\Lambda(M_1,M_2,M_3)\,.
\end{eqnarray}
\end{observation}
 
$\Lambda(M_1,M_2,M_3)$ has been classified in~\cite{szymanski2016classification}. Ours is a subcase where
$M_1,M_2,M_3$ are all real. From Fig. 1 of~\cite{szymanski2016classification}, the only possible shapes of the ruled surfaces on $\partial\Theta_+(H_1,H_2,H_3)$ is a cone shape.

\begin{figure}[htbp]
  \centering
  \includegraphics[width=2.5in]{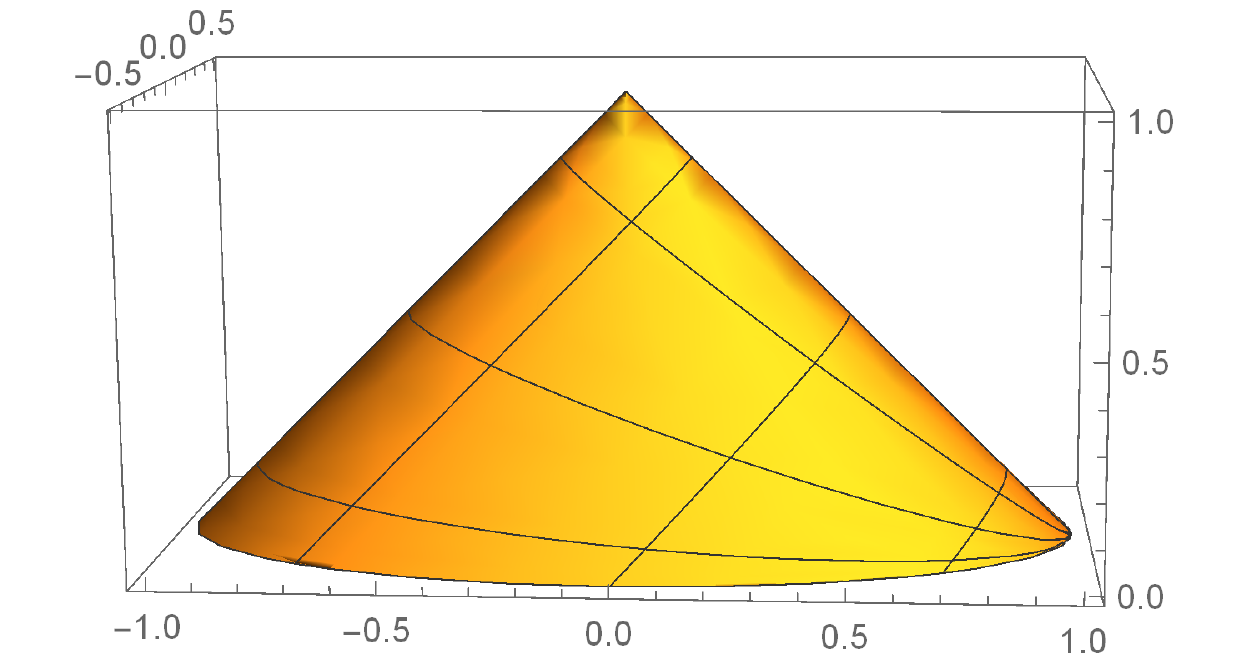}
  \caption{A cone shape $\Pi_+(H_1,H_2,H_3)$.}
  \label{fig:cone}
\end{figure}

A cone-shape $\Theta_+(H_1,H_2,H_3)$ can be given by block diagonal $M_i$\,s, for example $d_{xz,i}=d_{yz,i}=0$. For a concrete example, take
\begin{equation*}
M_1=
\begin{pmatrix}
1& 0 & 0\\
0 & -1 & 0\\
0& 0 & 0
\end{pmatrix},
M_2=
\begin{pmatrix}
0& 1 & 0\\
1 & 0 & 0\\
0& 0 & 0
\end{pmatrix},
M_3=
\begin{pmatrix}
0& 0 & 0\\
0 & 0 & 0\\
0& 0 & 1
\end{pmatrix}.
\end{equation*}
The corresponding Hamiltonian is given by
\begin{eqnarray}
H_1&=&X\otimes X-Y\otimes Y,\nonumber\\
H_2&=&X\otimes Y+Y\otimes X,\nonumber\\
H_3&=&Z\otimes Z\,.
\end{eqnarray}
The joint product numerical range $\Pi_+(H_1,H_2,H_3)$ has the form as shown in Fig.~\ref{fig:cone}.

\subsection{B. The non-homogenous case}

For the non-homogenous case (i.e. $f_i$s are not homogenous polynomials of $r,s,t$), we look at the two examples discussed in~\cite{chen2016physical}, which are many-body interacting boson systems in
the $N\rightarrow\infty$ limit.
The first example is the two model Ising model with
\begin{eqnarray}
H_1&=&X\otimes X\,,\nonumber\\
H_2&=&\frac{1}{2}(Z\otimes I+I\otimes Z)\,,\nonumber\\
H_3&=&\frac{1}{2}(X\otimes I+I\otimes X)\,.
\end{eqnarray}
This corresponds to
\begin{eqnarray}
f_1=r^2,\ f_2=t,\ f_3=r,
\end{eqnarray}
as formulated in Eq.~\eqref{eq:fi}.

\begin{figure}[htbp]
  \centering
  \centerline{\includegraphics[width=3.5in]{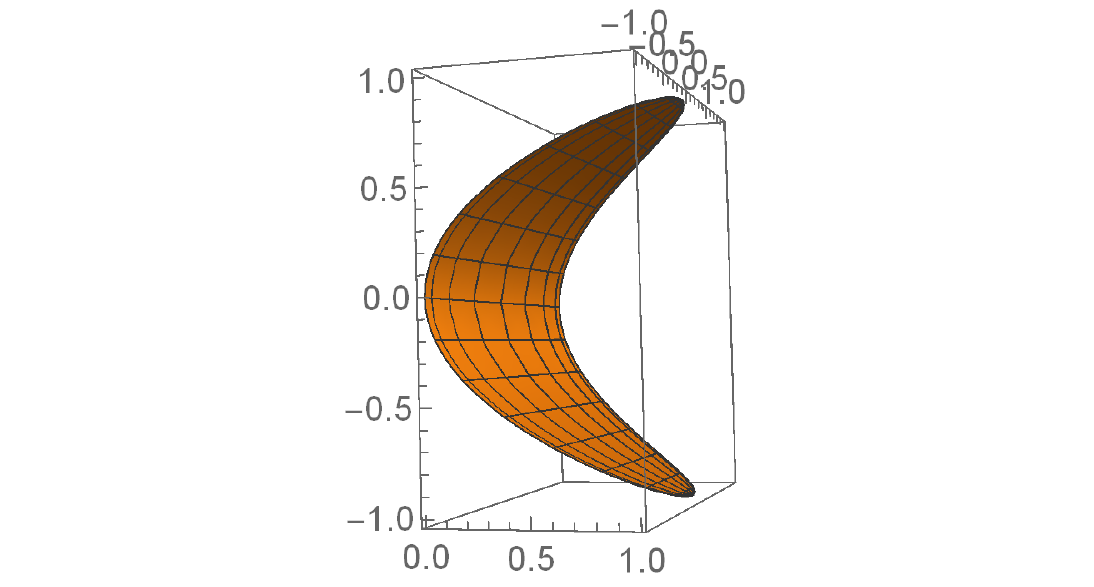}}
  \caption{$\Pi_+(H_1,H_2,H_3)$ for the two mode Ising model.}
  \label{fig:Ising}
\end{figure}

The joint product numerical range $\Pi_+(H_1,H_2,H_3)$ has the form as in Fig.~\ref{fig:Ising}.
This corresponds to the blue ruled surface of Fig.~1 in~\cite{chen2016physical}, which has a gapless physical origin. The green ruled surface of Fig.~1 in~\cite{chen2016physical} is a result of the convex hull of $\Pi_+(H_1,H_2,H_3)$, which corresponds to symmetry-breaking.

The second example in~\cite{chen2016physical} is the two model XY model with
\begin{eqnarray}
H_1&=&X\otimes X\,,\nonumber\\
H_2&=&Y\otimes Y\,,\nonumber\\
H_3&=&\frac{1}{2}(Z\otimes I+I\otimes Z)\,.
\end{eqnarray}
This corresponds to
\begin{eqnarray}
f_1=r^2,\ f_2=s^2,\ f_3=t,
\end{eqnarray}
as formulated in Eq.~\eqref{eq:fi}.

\begin{figure}[htbp]
  \centering
  \centerline{\includegraphics[width=3.5in]{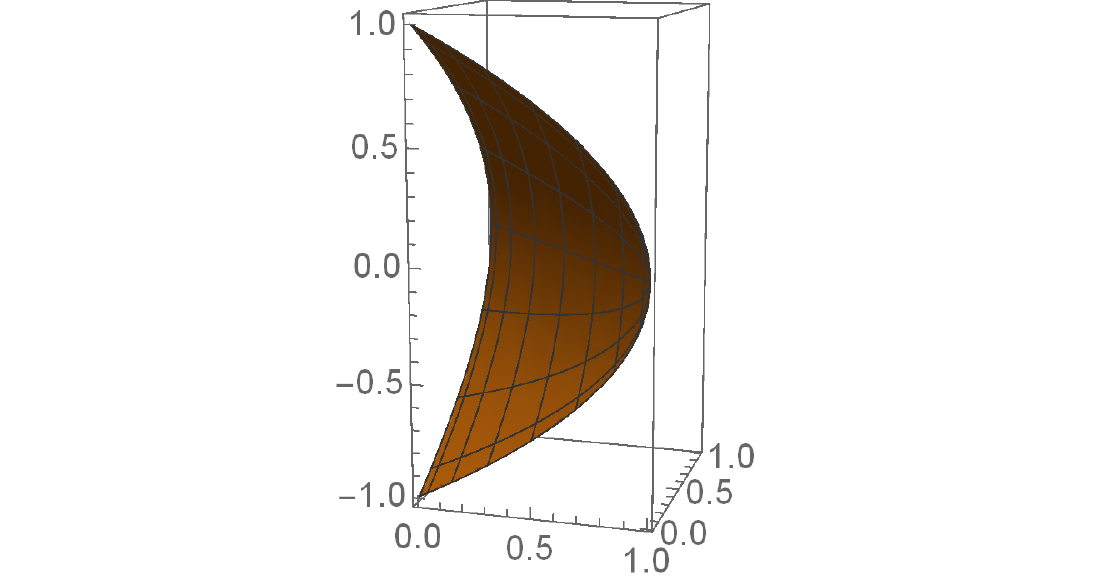}}
  \caption{$\Pi_+(H_1,H_2,H_3)$ for the two mode XY model.}
  \label{fig:XY}
\end{figure}

The joint product numerical range $\Pi_+(H_1,H_2,H_3)$ has the form as shown in Fig.~\ref{fig:XY}.
This corresponds to the blue ruled surface of Fig.~4 in~\cite{chen2016physical}, which has a gapless physical origin.

These examples support our idea that a ruled surface on $\partial\Theta_+(H_1,H_2,H_3)\cap\Pi_+(H_1,H_2,H_3)$ have a gapless origin, as discussed in
Sec.~IIIA.

\section{V. Summary and discussion}

In this work, we make a connection between joint product numerical range and reduced density matrix geometry. We focus on the case of systems in infinite spatial dimension, where the reduced density matrices are known to be separable due to the quantum de Finetti's theorem. In this scenario,  our main 
observation is that the intersection of the joint product numerical range $\Pi(H_1,H_2,H_3)/\Pi_+(H_1,H_2,H_3)$ with the boundary of its convex hull $\Theta(H_1,H_2,H_3)/\Theta_+(H_1,H_2,H_3)$ contains information on the physical properties of the system. In particularly, a ruled surface on $\partial\Theta(H_1,H_2,H_3)$ that is in $\Pi(H_1,H_2,H_3)$ has a gapless origin, otherwise it has a symmetry breaking origin.

Notice that it is possible to have the same $\Theta(H_1,H_2,H_3)/\Theta_+(H_1,H_2,H_3)$ that is the convex hull obtained from very different  $\Pi(H_1,H_2,H_3)\Pi_+(H_1,H_2,H_3)$. We provide some concrete examples in Appendix A. Similar ideas apply to
the case of the same $\Theta_+(H_1,H_2,H_3)$ with different $\Pi_+(H_1,H_2,H_3)$. Therefore, in practice, solely by looking at the shape of $\Theta(H_1,H_2,H_3)$
is not enough to tell the physical properties of the system. One will need to further
look at the shape of $\Pi(H_1,H_2,H_3)$, especially its intersection with
$\partial\Theta(H_1,H_2,H_3)$.

We provide a general method to obtain ruled surfaces on $\Theta(H_1,H_2,H_3)$, where $\Theta(H_1,H_2,H_3)$ is a convex hull of two convex objects. This allows us to construct $\Theta(H_1,H_2,H_3)$ of certain interesting geometric shapes, such as the oloid. Similar idea
is applied to the symmetric case where different shapes of ruled surfaces $\Theta(H_1,H_2,H_3)$
are obtained. 

It will be interesting to classify all possible shapes of 
the joint separable numerical range $\Theta(H_1,H_2,H_3)/\Theta_+(H_1,H_2,H_3)$, and the corresponding joint separable numerical range $\Pi(H_1,H_2,H_3)/\Pi_{+}(H_1,H_2,H_3)$, at least
in low (single-particle) dimensions, for systems with infinite spatial dimension where the quantum de Finetti's theorem is valid.
That will then contain information of all possible physical properties in these systems. 

\section*{Acknowledgement}

NY and BZ are supported by NSERC and CIFAR. This research was supported in part by Perimeter Institute for Theoretical Physics.
Research at Perimeter Institute is supported by the Government of Canada through Industry Canada and by the Province
of Ontario through the Ministry of Economic Development \& Innovation.

\appendix

\section{Appendix A. $\Theta(H_1,H_2,H_3)$ vs. $\Pi(H_1,H_2,H_3)$}

Let 
\[
X=\begin{pmatrix}
0 & 1   \\
1 & 0
\end{pmatrix},\ Y=\begin{pmatrix}
0 & -i   \\
i & 0
\end{pmatrix},\ Z=\begin{pmatrix}
1 & 0   \\
0 & -1
\end{pmatrix}\,.\]

The following three examples all have the same $\Theta_+(H_1,H_2,H_3)$. 

\medskip

\begin{figure}[hbpt]
  \centering
  \centerline{\includegraphics[width=2.5in]{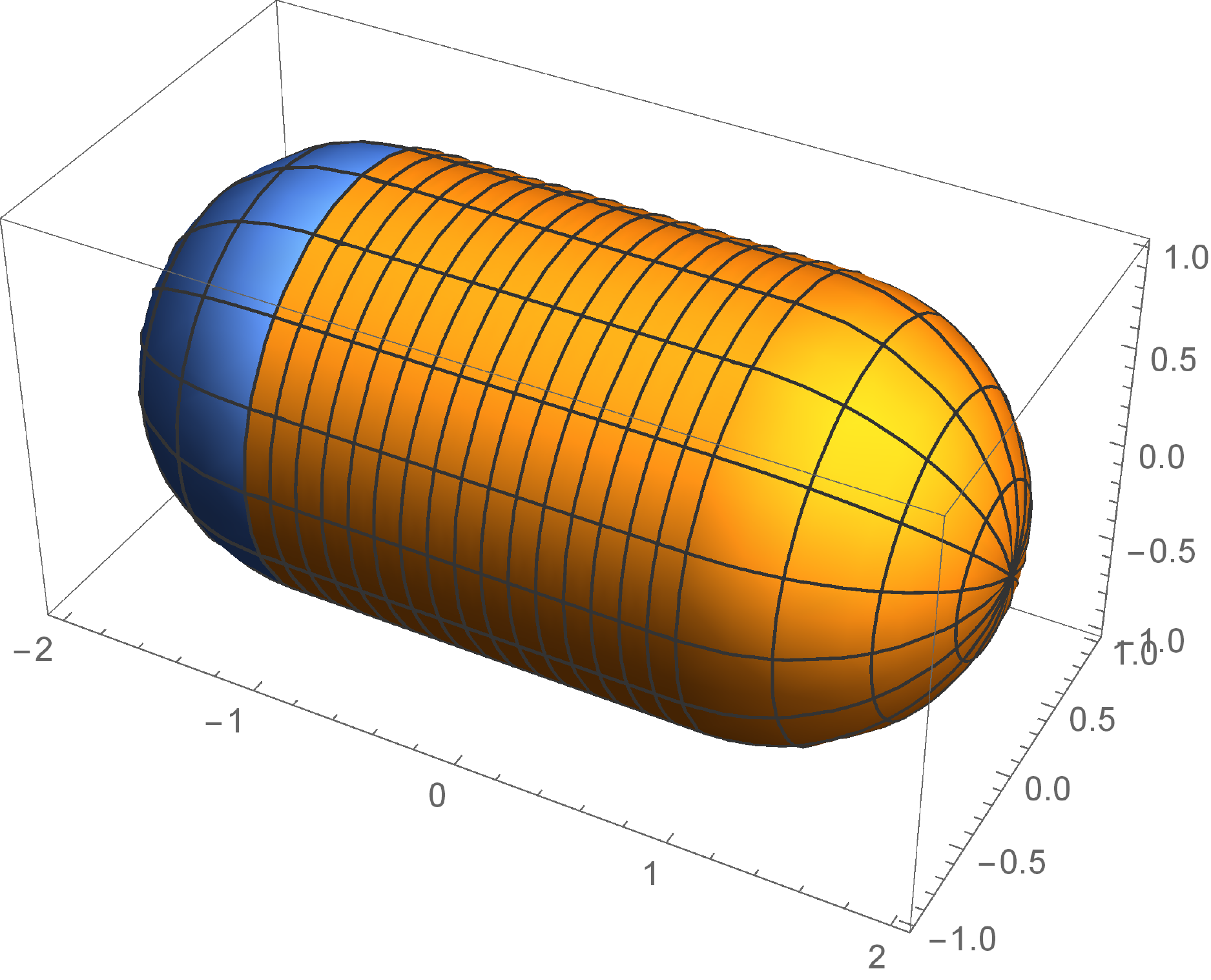}}
  \caption{$\Pi(H_1,H_2,H_3)$ for Example 1.}
  \label{fig:eg1}
\end{figure}

\noindent{\bf Example 1} Let $H_1=(I+Z)\oplus(-I+Z),\ H_2=X\oplus X$ and $H_3=Y\oplus Y$. Then
\begin{eqnarray*}
&&\Pi(H_1,H_2,H_3)=\nonumber\\
&&\{(2t-1+\cos \theta ,\sin \theta \cos\phi,\sin\theta\sin\phi):\nonumber\\
&&0\le t\le 1,\ 0\le \theta,\ \phi\le 2\pi\}\nonumber\\
&&=\Theta(H_1,H_2,H_3)\,.
\end{eqnarray*}
The corresponding $\Pi(H_1,H_2,H_3)$ is shown in Fig.~\ref{fig:eg1}.
Since the boundary ruled surface of $\Theta(H_1,H_2,H_3)$ is in 
$\Pi(H_1,H_2,H_3)$, it has a gapless origin as discussed in
Sec.~IIIA.

\medskip

\begin{figure}[htbp]
  \centering
\centerline{\includegraphics[width=2.5in]{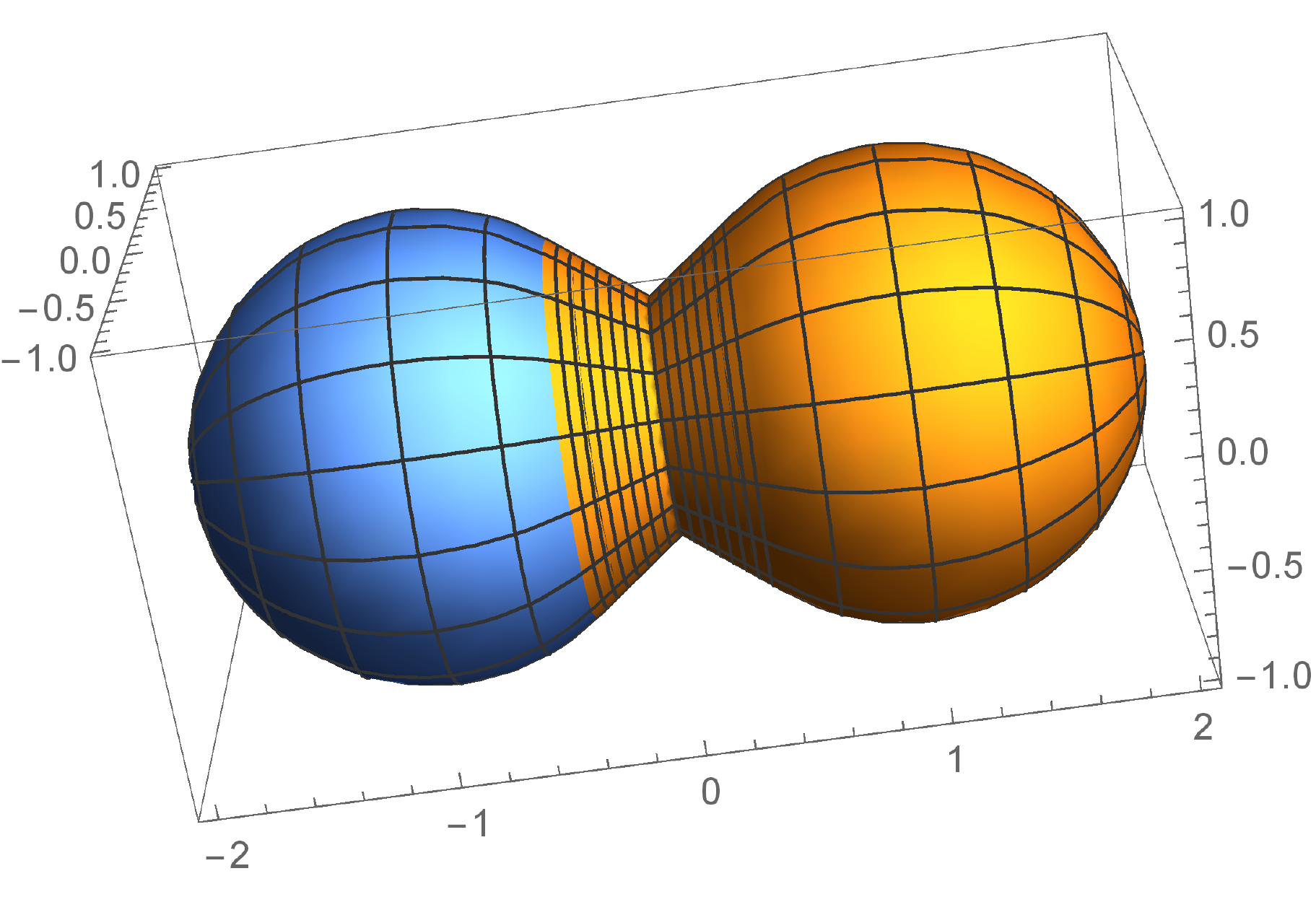}}
  \caption{$\Pi(H_1,H_2,H_3)$ for Example 2.}
  \label{fig:eg2}
\end{figure}

\noindent{\bf Example 2} Let $H_1=(I+Z)\oplus(-I+Z),\ H_2=X\oplus(- X)$ and $H_3=Y\oplus( -Y)$. Then
\begin{eqnarray*}
&&\Pi(H_1,H_2,H_3)=\nonumber\\
&&\{(2t-1+\cos \theta ,(2t-1)\sin \theta \cos\phi,(2t-1)\sin\theta\sin\phi):\nonumber\\
&&0\le t\le 1,\ 0\le \theta,\ \phi\le 2\pi\}\,.
\end{eqnarray*}
The corresponding $\Pi(H_1,H_2,H_3)$ is shown in Fig.~\ref{fig:eg2}.
Since the boundary ruled surface of $\Theta(H_1,H_2,H_3)$ is not in 
$\Pi(H_1,H_2,H_3)$, it has a symmetry breaking origin as discussed in
Sec.~IIIA.

\medskip

\begin{figure}[hbpt]
  \centering
\centerline{\includegraphics[width=2.5in]{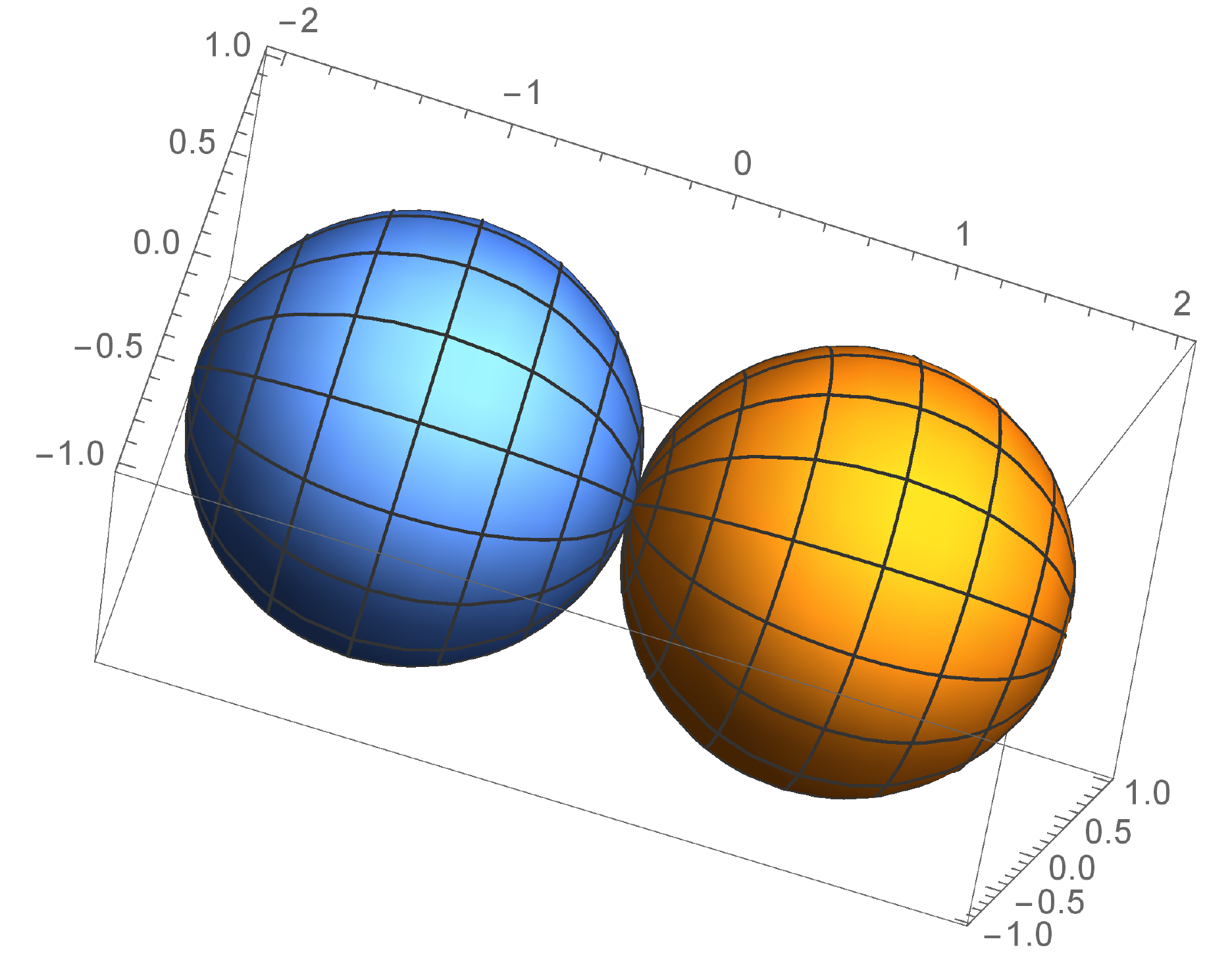}}
  \caption{$\Pi(H_1,H_2,H_3)$ for Example 3.}
  \label{fig:eg3}
\end{figure}

\noindent{\bf Example 3} Let $H_1=(I+Z)\oplus(-I-Z),\ H_2=X\oplus(- X)$ and $H_3=Y\oplus( -Y)$. Then
\begin{eqnarray*}
&&\Pi(H_1,H_2,H_3)=\nonumber\\
&&\{((2t-1)\cos \theta ,(2t-1)\sin \theta \cos\phi,(2t-1)\sin\theta\sin\phi):\nonumber\\
&&0\le t\le 1,\ 0\le \theta,\ \phi\le 2\pi\}\,.
\end{eqnarray*}
The corresponding $\Pi(H_1,H_2,H_3)$ is shown in Fig.~\ref{fig:eg3}.
Since the boundary ruled surface of $\Theta(H_1,H_2,H_3)$ is not in 
$\Pi(H_1,H_2,H_3)$, similar to Example 2, it has a symmetry breaking origin.

\bibliography{ProNR}

\end{document}